\documentclass{desyprocA4}

\def\to{\rightarrow}

\def\bi{\begin{itemize}}
 \def\ei{\end{itemize}}

\def\c1p{C1^\prime}
\def\msq3{\overline{m}_{\tilde{q}}(3)}
\def\ta{\tilde a}

\def\ta{\tilde a}

\def\tst{\tilde t}

\def\tg{\tilde g}

\def\be{\begin{equation}}  
\def\ee{\end{equation}}  
\def\bea{\begin{eqnarray}}  
\def\eea{\end{eqnarray}}

\def\tz{\widetilde Z}

\def\beq{\begin{equation}}
\def\eeq#1{\label{#1}\end{equation}}
\def\eeqn{\end{equation}}

\newenvironment{Eqnarray}%
   {\arraycolsep 0.14em\begin{eqnarray}}{\end{eqnarray}}
\def\beqa{\begin{Eqnarray}}
\def\eeqa#1{\label{#1}\end{Eqnarray}}
\def\eeqan{\end{Eqnarray}}

\begin{document}
\title{Snowmass whitepaper: Exploring natural SUSY via \\
direct and indirect detection of higgsino-like WIMPs
}

\author{{\slshape H. Baer$^1$, V. Barger$^2$, D. Mickelson$^1$ and X. Tata$^3$}\\
$^1$Dept. of Physics and Astronomy, University of Oklahoma, Norman, OK 73019, USA\\ 
$^2$Dept. of Physics, University of Wisconsin, Madison, WI 53706, USA\\
$^3$Dept. of Physics and Astronomy, University of Hawaii, Honolulu, HI 96822, USA\\
}


\maketitle


\begin{abstract}
Supersymmetric models with low electroweak fine-tuning contain light
higgsinos with mass not too far from $m_h\simeq 125$ GeV, while other
sparticles can be much heavier.  In the $R$-parity conserving MSSM, the
lightest neutralino is then a higgsino-like WIMP (albeit with
non-negligible gaugino components), with thermal relic density well
below measured values. This leaves room for axions (or other, perhaps
not as well motivated, stable particles) to function as co-dark matter
particles. The local WIMP abundance is then expected to be below
standard estimates, and direct and indirect detection rates must be
accordingly rescaled. We calculate rescaled direct and indirect
higgsino-like WIMP detection rates in SUSY models that fulfil the
electroweak naturalness condition. In spite of the rescaling, we find
that ton-scale noble liquid detectors can probe the entire higgsino-like
WIMP parameter space, so that these experiments should either discover
WIMPs or exclude the concept of electroweak naturalness in $R$-parity
conserving natural SUSY models.  Prospects for spin-dependent or
indirect detection are more limited due in part to the rescaling effect.
\end{abstract}

%

In previous studies \cite{ltr,rns,white1} it has been argued that 
a necessary condition for naturalness of SUSY models is the requirement of no
large uncorrelated cancellations to $m_Z^2/2$ in the one-loop effective potential
minimization condition
\be \frac{m_Z^2}{2} =
\frac{m_{H_d}^2 + \Sigma_d^d -
(m_{H_u}^2+\Sigma_u^u)\tan^2\beta}{\tan^2\beta -1} -\mu^2 \;.
\label{eq:loopmin}
\ee 
Here, Eq.~(\ref{eq:loopmin}) is implemented as a {\it weak scale
relation}, even for SUSY theories purporting to be valid all the way up
to scales as high as $M_{\rm GUT}-M_{P}$.  The quantities $\Sigma_u^u$
and $\Sigma_d^d$ are the one-loop corrections arising from loops of
particles and their superpartners that couple directly to the Higgs
doublets. Thus, electroweak naturalness requires 1. low $\mu\sim
100-300$ GeV, 2. $m_{H_u}^2$ is driven to small negative values so that
$m_{H_u}^2({\rm weak})\sim 100-300$ GeV and 3. highly mixed TeV-scale top
squarks. The large mixing both reduces the $\tst_1$ and $\tst_2$
contributions to $\Sigma_u^u$ while lifting the Higgs mass $m_h\sim 125$
GeV. Models where these conditions are met have a low value of
$\Delta_{EW}$ \cite{ltr,rns,white1} and have been labeled as
radiatively-driven natural supersymmetry or RNS.

\begin{figure}[tbp]
\begin{center}
\includegraphics[height=0.3\textheight]{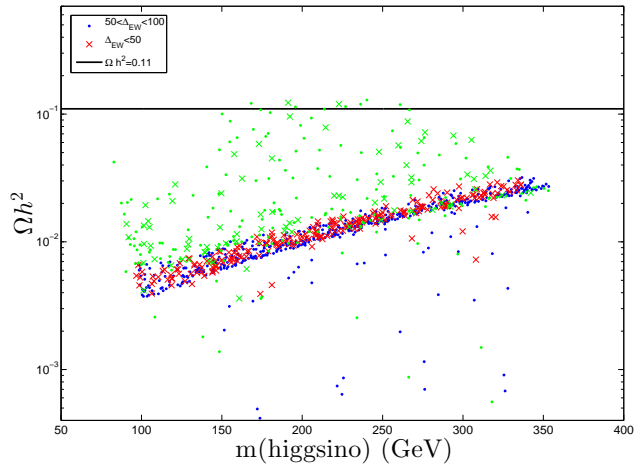}
\caption{Plot of standard thermal neutralino abundance 
$\Omega_{\tz_1}^{std}h^2$ versus $m(higgsino)$ 
from a scan over NUHM2 parameter space with $\Delta_{EW}<50$ (red crosses) and $\Delta_{EW}<100$ (blue dots). 
Green points are excluded by current direct/indirect WIMP search experiments.
We also show the central value of $\Omega_{CDM}h^2$ from WMAP9.
\label{fig:oh2}}
\end{center}
\end{figure}

In RNS models, since $\mu\sim 100-300$~GeV, then the lightest neutralino
is largely higgsino-like but with a non-negligible gaugino
component.\footnote{ In the RNS model, if gaugino mass $M_3$ is too
large, it lifts the top squark masses high enough so that the radiative
corrections $\Sigma_u^u(\tst_{1,2})$ become large leading to
fine-tuning.  Since the gaugino masses $M_1$ and $M_2$ are related to
$M_3$ under gaugino mass unification, then the lightest neutralino--
while dominantly higgsino-like-- always has a non-negligible gaugino
component. In models with light higgsinos but with very large gaugino masses
(such as Br\"ummer-Buchm\"uller model\cite{bb}), the lightest neutralino
is nearly pure higgsino.  In that case, the spin-independent direct
detection rates presented below {\it will not obtain} since the
$h-\tz_1\tz_1$ coupling depends on a product of higgsino and gaugino
components\cite{bbm}.}  The thermally-produced (TP) relic density of
higgsino-like WIMPs from RNS has been calculated in
Ref's~\cite{rns,bbm}; it is typically found to be a factor 5-15 below
measured value for CDM,  as shown in Fig. \ref{fig:oh2}.  To accommodate this
situation, a cosmology with mixed axion/higgsino dark matter (two dark
matter particles, an axion and a higgsino-like neutralino)\cite{ckls}
has been invoked in Ref.~\cite{rns,bbm}.  In this case, thermal
production of axinos $\ta$ in the early universe followed by $\ta\to
g\tg,\ \gamma\tz_i$ leads to additional neutralino production. In the
case where axinos are sufficiently produced, their decays may lead to
neutralino re-annihilation at temperatures below freeze-out; the
resulting re-annihilation abundance is always larger than the standard
freeze-out value.  In addition, coherent-oscillation production of
saxions $s$ at high PQ scale $f_a>10^{12}$~GeV followed by saxion decays
to SUSY particles can also augment the neutralino abundance. Late saxion
decay to primarily SM particles can result in entropy dilution of all
relics (including axions) present at the time of decay, so long as BBN
and dark radiation constraints are respected.  The upshot is that,
depending on the additional PQ parameters, either higgsino-like
neutralinos or axions can dominate the dark matter abundance, or they
may co-exist with comparable abundances: this leads to the possibility
of detecting both an axion and a WIMP.

\begin{figure}[tbp]
\begin{center}
\includegraphics[height=0.3\textheight]{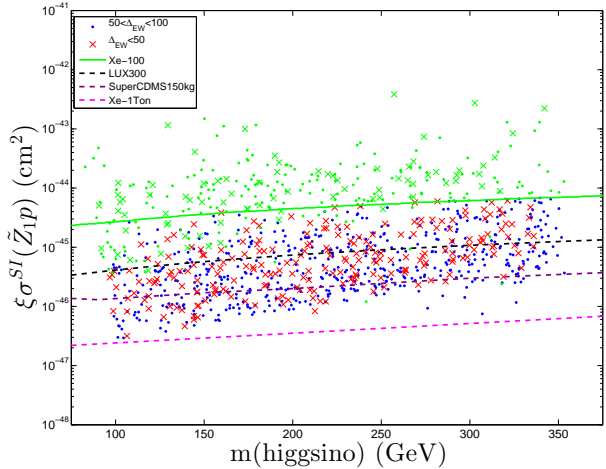}
\caption{Plot of rescaled higgsino-like WIMP spin-independent
direct detection rate $\xi \sigma^{SI}(\tz_1 p)$
versus $m(higgsino)$ from a scan over NUHM2 parameter space with $\Delta_{EW}<50$ (red crosses)
and $\Delta_{EW}<100$ (blue dots).
Green points are excluded by current direct/indirect WIMP search experiments.
We also show the current limit from $Xe$-100 experiment,
and projected reaches of LUX, SuperCDMS 150 kg and $Xe$-1 ton.
\label{fig:SI}}
\end{center}
\end{figure}

In the case of mixed axion-WIMP dark matter, the local WIMP abundance
might be well below the commonly accepted local abundance
$\rho_{loc}\sim 0.3$~GeV/cm$^3$. Thus, to be conservative, limits from
experiments like Xe-100 or CDMS should be compared to theoretical
predictions which have been scaled down\cite{bottino} by a factor
$\xi\equiv \Omega_{\tilde h}^{TP}h^2/0.12$.  In the RNS model, the
gauginos cannot be too heavy so that the neutralino always has a
substantial gaugino component even though it is primarily higgsino.
This means that spin-independent direct detection rates
$\sigma_{SI}(\tz_1p)$ are never too small.  Predictions for
spin-independent higgsino-proton scattering cross section are shown in
Fig. \ref{fig:SI} and compared against current limits and future reach
projections.  
Even accounting for the local scaling factor $\xi$, it is
found\cite{bbm} that ton-scale noble liquid detectors such as
Xe-1-ton\cite{xe1ton} should completely probe the model parameter
space. One caveat is that if saxions give rise to huge entropy dilution
after freeze-out while avoiding constraints from dark radiation and BBN,
then the local abundance may be even lower than the assumed freeze-out
value, and the dark matter would be highly axion dominated.

\begin{figure}[tbp]
\includegraphics[height=0.25\textheight]{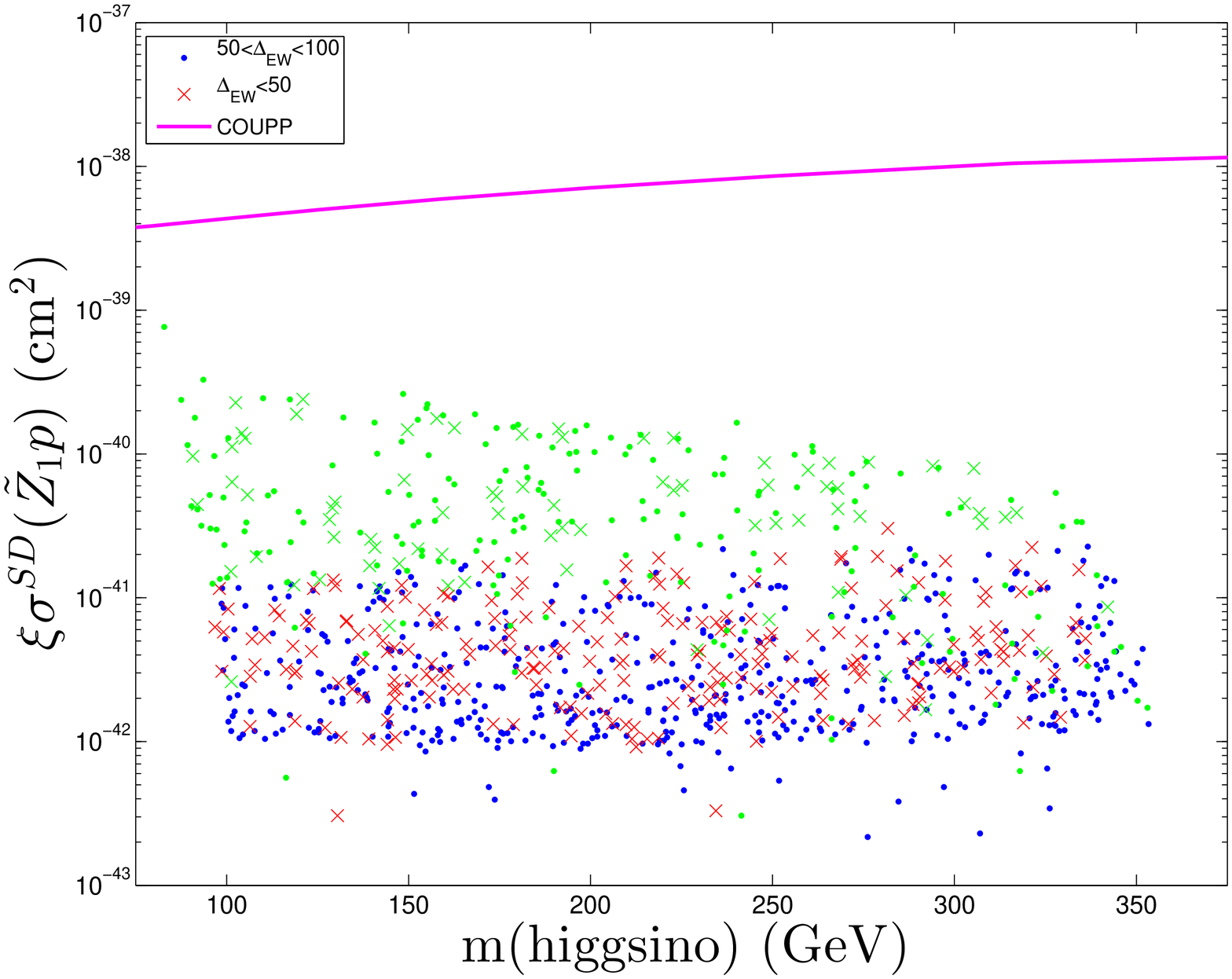}
\includegraphics[height=0.25\textheight]{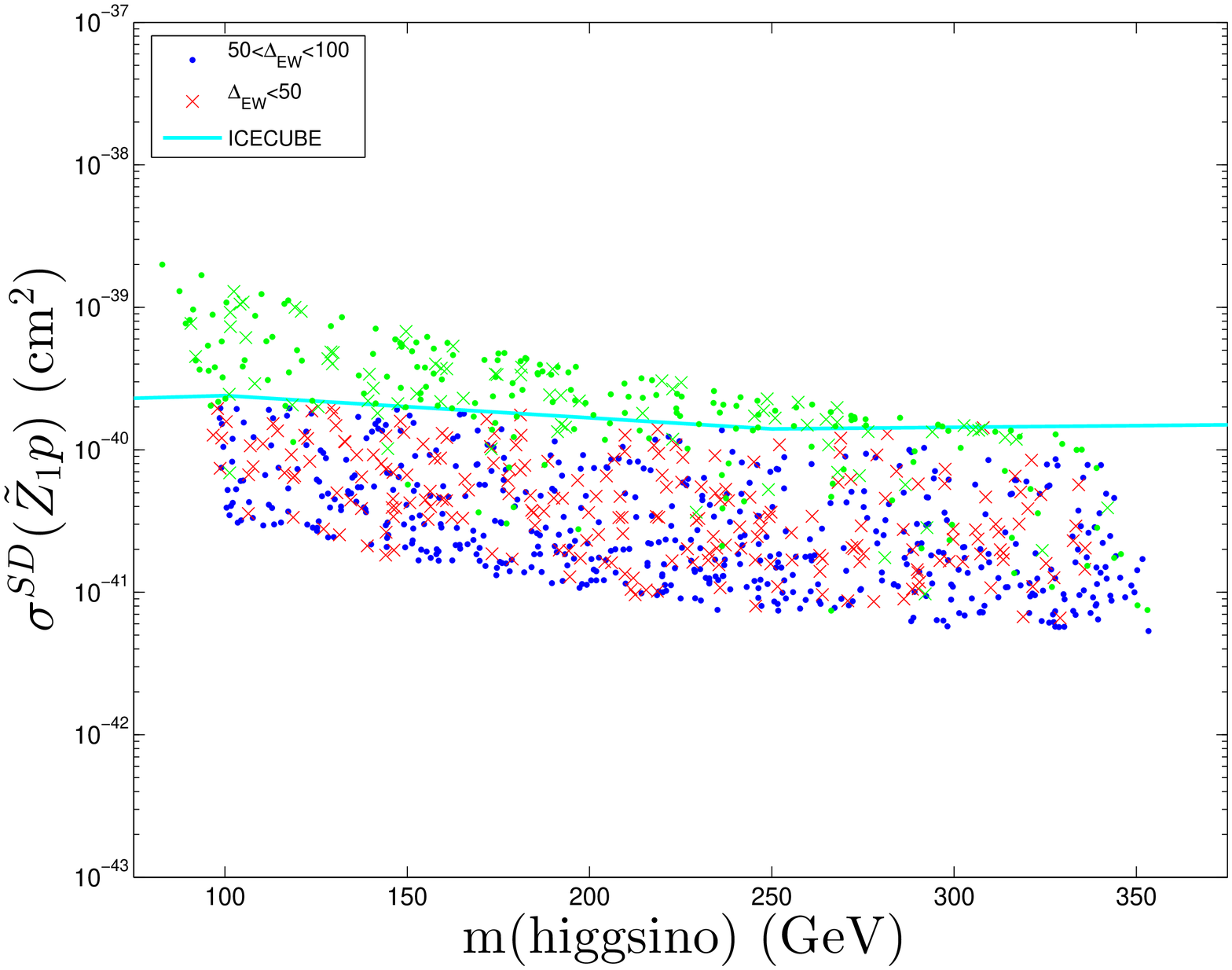}
\caption{In {\it a}), we plot the rescaled spin-dependent higgsino-like WIMP detection rate 
$\xi \sigma^{SD}(\tz_1 p)$ versus $m(higgsino)$ from a scan over NUHM2 parameter space with $\Delta_{EW}<50$ (red crosses)
and $\Delta_{EW}<100$ (blue dots). 
Green points are excluded by current direct/indirect WIMP search experiments.
The curve shows the current limit from the COUPP experiment, Ref.~\cite{coupp}.
In {\it b}), we plot the (non-rescaled) spin-dependent higgsino-like WIMP detection rate $\sigma^{SD}(\tz_1 p)$. 
The curve shows the current limit from the IceCube experiment, Ref.~\cite{icecube}.
\label{fig:SD}}
\end{figure}

In Fig. \ref{fig:SD}{\it a}), we show the rescaled {\it spin-dependent}
higgsino-proton scattering cross section $\xi\sigma^{SD}(\tz_1 p)$. Here
we show recent limits from the COUPP\cite{coupp} detector. Current
limits are still about an order of magnitude away from reaching the
predicted rates from RNS models.
To compare against the current reach of IceCube\cite{icecube}, we show in Fig. \ref{fig:SD}{\it b})
the value of $\sigma^{SD}(\tz_1 p)$ but with no rescaling factor. 
Here,  the IceCube rates should not be rescaled since the IceCube detection depends on whether the Sun has 
equilibrated its core abundance between capture rate and annihilation rate\cite{gjk}. Typically for the Sun, 
equilibration is reached for almost all of SUSY parameter space\cite{bottino_nu}.
The IceCube limits have entered the RNS parameter space and excluded the largest values
of $\sigma^{SD}(\tz_1 p)$.  

\begin{figure}[tbp]
\begin{center}
\includegraphics[height=0.29\textheight]{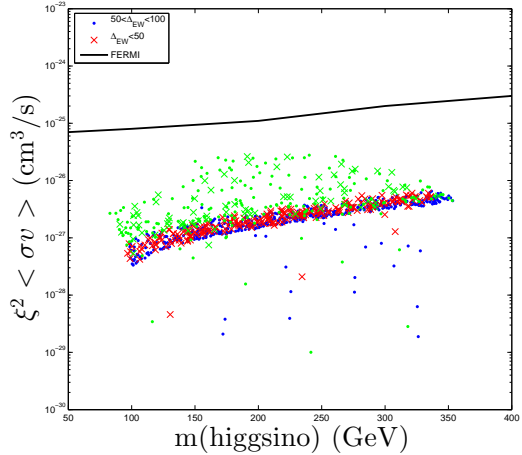}
\caption{Plot of rescaled $\xi^2 \langle\sigma v\rangle |_{v\to 0}$ 
versus $m(higgsino)$ from a scan over NUHM2 parameter space with $\Delta_{EW}<50$ (red crosses)
and $\Delta_{EW}<100$ (blue dots). 
Green points are excluded by current direct/indirect WIMP search experiments.
The curve shows the current limit from Fermi LAT, Ref.~\cite{fermi}.
\label{fig:sigv}}
\end{center}
\end{figure}

In Fig.~\ref{fig:sigv}, we show the rescaled thermally-averaged
neutralino annihilation cross section times relative velocity in the
limit as $v\to 0$: $\xi^2\langle\sigma v\rangle|_{v\to 0}$.  This
quantity enters into the rate expected from WIMP halo annihilations into
$\gamma$, $e^+$, $\bar{p}$ or $\overline{D}$.  The rescaling appears as
$\xi^2$ since limits depend on the square of the local WIMP
abundance\cite{bottino_id}.  Anomalies in the positron and $\gamma$
spectra have been reported, although the former may be attributed to
pulsars\cite{pulsars}, while the latter 130 GeV gamma line may be
instrumental.  Soon to be released results from AMS-02 should clarify
the situation\cite{Aguilar:2013qda}. On the plot, we show the limit
derived from the Fermi LAT gamma ray observatory\cite{fermi} for WIMP
annihilations into $WW$.  These limits have not yet reached the RNS
parameter space due in part to the squared rescaling factor.

\section{Acknowledgments} This research was sponsored in part by grants from the US Department of Energy


\begin{footnotesize}

\end{footnotesize}



\begin{thebibliography}{99}
%
\bibitem{ltr} H.~Baer, V.~Barger, P.~Huang, A. Mustafayev and X.~Tata,
  Phys. Rev. Lett. {\bf 109}, 161802 (2012).
%
\bibitem{rns} H.~Baer, V.~Barger, P.~Huang, D. Mickelson,
A. Mustafayev and X.~Tata, arXiv:1212.2655 (2012) (PRD, in press).
%
\bibitem{white1} H.~Baer, V.~Barger, P.~Huang, D. Mickelson,
A. Mustafayev and X.~Tata, arXiv:1306.2926 (2013).
%
\bibitem{bb}
  F.~Brummer and W.~Buchmuller,
  JHEP {\bf 1205} (2012) 006.
%
\bibitem{bbm} H.~Baer, V.~Barger and D.~Mickelson,
  arXiv:1303.3816 [hep-ph].
%
\bibitem{ckls}  K.~-Y.~Choi, J.~E.~Kim, H.~M.~Lee and O.~Seto,
  Phys.\ Rev.\ D {\bf 77} (2008) 123501;
H. Baer, A. Lessa, S. Rajagopalan and W. Sreethawong, JCAP {\bf 1106} (2011) 031; 
H. Baer, A. Lessa and W. Sreethawong, JCAP {\bf 1201} (2012) 036;
K.~J.~Bae, H.~Baer and A.~Lessa,
  JCAP {\bf 1304} (2013) 041.
%
\bibitem{bottino} A.~Bottino, F.~Donato, N.~Fornengo and S.~Scopel,
  Phys.\ Rev.\ D {\bf 63} (2001) 125003.
%
\bibitem{xe1ton} E.~Aprile [XENON1T Collaboration],
  arXiv:1206.6288.
%
\bibitem{coupp} E.~Behnke {\it et al.}  [COUPP Collaboration],
  Phys.\ Rev.\ D {\bf 86} (2012) 052001.
%
\bibitem{icecube}  R.~Abbasi {\it et al.} (IceCube collaboration),
   Phys. Rev. {\bf D85} (2012) 042002.
%
\bibitem{gjk} G.~Jungman, M.~Kamionkowski and K.~Griest,
  Phys.\ Rept.\  {\bf 267} (1996) 195.
%
\bibitem{bottino_nu} V.~Niro, A.~Bottino, N.~Fornengo and S.~Scopel,
  Phys.\ Rev.\ D {\bf 80} (2009) 095019.
%
\bibitem{bottino_id} A.~Bottino, F.~Donato, N.~Fornengo and P.~Salati,
  Phys.\ Rev.\ D {\bf 72} (2005) 083518.
%
\bibitem{pulsars} V.~Barger, Y.~Gao, W.~Y.~Keung, D.~Marfatia and G.~Shaughnessy,
  Phys.\ Lett.\ B {\bf 678} (2009) 283 ;
S.~Profumo,
  Central Eur.\ J.\ Phys.\  {\bf 10} (2011) 1.

%
\bibitem{fermi}  M.~Ackermann {\em et al.} (Fermi Collaboration), 
Phys. Rev. Lett. {\bf 107} (2011) 241302;
A.~Geringer-Sameth and S.~M.~Koushiappas, 
Phys. Rev. Lett. {\bf 107} (2011) 241303.
%
\bibitem{Aguilar:2013qda}
  M.~Aguilar {\it et al.}  [AMS Collaboration],
  Phys.\ Rev.\ Lett.\  {\bf 110} (2013) 14,  141102.

\end{thebibliography}
\end{document}